# Planetary Trojans – the main source of short period comets?

J. Horner[1] & P. S. Lykawka[2]

[1] Department of Physics, Science Laboratories, University of Durham, South Road, Durham, UK, DH1 3LE

[2] Astronomy Group, Faculty of Social and Natural Sciences, Kinki University, Shinkamikosaka 228-3, Higashiosaka-shi, Osaka, 577-0813, Japan

**Abstract**

One of the key considerations when assessing the potential habitability of telluric worlds will be that of the impact regime experienced by the planet. In this work, we present a short review of our understanding of the impact regime experienced by the terrestrial planets within our own Solar system, describing the three populations of potentially hazardous objects which move on orbits that take them through the inner Solar system. Of these populations, the origins of two (the Near-Earth Asteroids and the Long-Period Comets) are well understood, with members originating in the Asteroid belt and Oort cloud, respectively. By contrast, the source of the third population, the Short-Period Comets, is still under debate. The proximate source of these objects is the Centaurs, a population of dynamically unstable objects that pass perihelion (closest approach to the Sun) between the orbits of Jupiter and Neptune. However, a variety of different origins have been suggested for the Centaur population. Here, we present evidence that at least a significant fraction of the Centaur population can be sourced from the planetary Trojan clouds, stable reservoirs of objects moving in 1:1 mean-motion resonance with the giant planets (primarily Jupiter and Neptune). Focussing on simulations of the Neptunian Trojan population, we show that an ongoing flux of objects should be leaving that region to move on orbits within the Centaur population. With conservative estimates of the flux from the Neptunian Trojan clouds, we show that their contribution to that population could be of order ~3%, while more realistic estimates suggest that the Neptune Trojans could even be the main source of fresh Centaurs. We suggest that further observational work is needed to constrain the contribution made by the Neptune Trojans to the ongoing flux of material to the inner Solar system, and believe that future studies of the habitability of exoplanetary systems should take care not to neglect the contribution of resonant objects (such as planetary Trojans) to the impact flux that could be experienced by potentially habitable worlds.

# 1. Introduction

Everywhere we look within our Solar system, we see evidence of impacts. Mercury is a pockmarked husk, the small remnant of a once larger differentiated planet (e.g. Benz et al. 2008), and revealed by images from passing spacecraft to be covered in innumerable craters, young and old. Despite Venus's thick atmosphere, which shields the planet's surface from small impactors, rapidly removes the evidence of impact scars, and makes it difficult to detect such scars in the first place, numerous impact features are known on the surface of Venus. The Moon, our nearest neighbour, is covered in the evidence of impacts both recent and dating back as far as the putative Late Heavy Bombardment (e.g. Gomes et al., 2005). Mars, too, bears the scars of innumerable impacts, from the giant basin which covers the entire northern hemisphere of the planet (Andrews-Hanna et al. 2008) to the myriad smaller craters imaged by orbiting craft (e.g. Grant et al., 2008) and visited by roving vehicles (e.g. Cabrol et al., 2006; Calvin, W. M. et al., 2008).

Compared with Mercury, the Moon and Mars, relatively few impact craters are known on the Earth. The current tally stands at just 176[1] confirmed impact features, a value which reflects the efficient weathering which removes such features from the surface on geological timescales, the difficulty in detecting and confirming such features (which are particularly well hidden by e.g. rainforests, oceans, and polar caps), and the fact that impactors are very efficiently slowed upon hitting the ocean, such that only the largest can leave craters beneath the ~70% of our planet which is covered by the oceans (Baldwin et al., 2007; Milner et al., 2008).

Despite the relative paucity of impact structures on the Earth, it is clear that impacts continue to rain down upon our planet. Such impacts can have significant effects on the biosphere of the planet, and have been proposed as the cause of a number of mass extinctions through the Earth's history. The extinction event which has drawn most attention over the years was that which involved the death of the dinosaurs, 65 million years ago. Many astronomers believe that the major factor in that extinction was the impact of a large (~10km) asteroidal or cometary body, creating the Chicxulub impact crater (e.g. Brett, 1992; Bottke et al., 2007), though a number of other mechanisms are considered more likely by scientists in other fields (e.g. Glasby & Kunzendorf, 1996; Poinar & Poinar, 2008). Regardless of what truly killed the dinosaurs, however, the idea that they *could* have been wiped out by the impact of a giant rock from space highlights the importance of understanding the processes by which such objects can be placed on Earth-encountering orbits. Such understanding is also vitally important when it comes to considering the search for life beyond our Solar system as is discussed in the review by Horner & Jones (2010), elsewhere in this proceedings.

Within our Solar system, there are three distinct groups of objects which contribute to the impact flux at Earth. The near-Earth asteroids, rocky and metallic bodies ranging in size up to the 32km diameter Ganymed, are currently believed to be the main contributors to the terrestrial impact flux, with smaller contributions coming from short- and long-period comets (e.g. Chapman 1994; Bottke et al., 2002). The three populations of potentially threatening objects are all dynamically unstable, and would be expected to become depleted on timescales of, at most, a few million years, if they were not continually resupplied from reservoirs of more stable parent objects.

The near-Earth asteroids are sourced from the asteroid belt (e.g. Morbidelli et al. 2002), and the mechanisms by which they are delivered to the Earth are believed to be well understood (as discussed in that work). The long-period comets are sourced from the Oort-cloud, a vast collection of cometary nuclei held in cold storage in a cloud believed to stretch half-way to the nearest star (e.g. Oort, 1950), and are thought to be injected to the inner Solar system as a result of perturbations on the Oort-cloud by passing stars (e.g. Oort, 1950; Biermann & Huebner, 1983), giant molecular clouds (e.g. Mazeeva, 2004; Thaddeus & Chanan, 1985), the galactic tide (e.g. Heisler & Tremaine,

---

[1] According to the Earth Impact Database, at http://www.unb.ca/passc/ImpactDatabase/, accessed 23rd April 2010

1986; Fouchard et al., 2005), and even, potentially, a massive companion to the Sun (e.g. Horner & Evans, 2002; Matese et al., 1999; Murray, 1999; Matese & Whitmire, 2010).

The origin of the short-period comets, however, is still the subject of some debate. While the proximate source of these objects is well known to be the Centaurs (e.g. Horner et al., 2003, 2004a, 2004b; Tiscareno & Malhotra 2003; di Sisto & Brunini 2007), the origin of the Centaurs themselves is still not well constrained. Initially, the Edgeworth-Kuiper belt[2] was mooted as the main source region for the Centaurs (e.g. Levison & Duncan, 1997). Indeed, the belt itself was independently predicted by both Edgeworth (1943) and Kuiper (1951) to explain the short-period comet population, long before the first members were discovered. Kuiper, however, suggested that such a belt would likely no longer be present as a result of dynamical scattering by Pluto, which was then still believed to be a planet-mass object. More recent studies have suggested that the key source of Centaurs is either the Scattered Disk[3] (e.g. Volk, K. & Malhotra, M., 2008) or the inner Oort Cloud (e.g. Emel'Yanenko et al., 2005, 2007). Horner & Evans (2006) also showed that some Centaurs can be captured into 1:1 mean-motion resonance with the giant planets (primarily Jupiter), becoming temporary members of their Trojan populations (in a solely gravitational set up consisting of the Sun and the host planet (and possibly other objects in distant orbits), any route by which an object can be dynamically captured can also be followed allowing an object to escape to non-Trojan space). Given that effects such as dynamical capture are time-reversible processes, the authors suggested that the Jovian Trojan population might act as an additional source of material to the Centaur population, contributing some small fraction of the total flux of fresh cometary material into the inner Solar system.

In this work, we propose a second resonant source of material for the Centaurs, namely the Neptune Trojan population. In section 2, we provide a brief description of this recently discovered addition to the menagerie of Solar system objects, before describing briefly a variety of dynamical studies we have carried out into both their formation and the evolution of current members of the family. In section 3, we provide exemplar results showing how Neptune Trojans can evolve to become short-period comets, before detailing simple calculations which suggest the Neptune Trojans might even represent the main source of material to the Centaur population. Finally, in section four, we discuss the implications of our work for our understanding of habitability in our own Solar system and beyond, and draw our conclusions.

---

[2] The Edgeworth-Kuiper belt is a disk of objects beyond the orbit of Neptune, most of which move on orbits that are dynamically stable on very long timescales. The inner edge of the Edgeworth-Kuiper belt is generally accepted to lie at the location of the Neptunian 2:3 MMR, at 39.5 AU, and the outer edge lies near the 1:2 MMR with the same planet (~48 AU). Objects moving within this region on dynamically stable orbits are considered members of the Edgeworth-Kuiper belt (often referred to as the Classical belt, or Kuiper belt). Although it is composed of icy, rather than rocky bodies, and likely contains significantly more mass, and more objects (both large and small) than the asteroid belt, it is in many ways analogous to that reservoir. The orbits of objects within the Edgeworth-Kuiper belt are spread over a wide range of inclinations, but are of typically low eccentricity – just as is the case for objects in the Asteroid belt.

[3] The Scattered Disk is a population of objects beyond the orbit of Neptune which undoubtedly bears close ties with the Edgeworth-Kuiper belt. Scattered Disk objects move on orbits with greater eccentricities than those in the Edgeworth-Kuiper belt, with many having perihelia closer to the orbit of Neptune than the objects in that belt. While the majority of objects in the Edgeworth-Kuiper belt are dynamically stable, however, those in the Scattered Disk are not – over time, their orbits can be perturbed by the distant influence of the massive planets until they become Neptune-crossing objects. Many Scattered Disk objects move on orbits so eccentric their aphelia lie hundreds, or even thousands of AU from the Sun – and it is likely that there is some overlap between that population and the inner-Oort cloud. It has been suggested that the Scattered Disk is the dynamically unstable counterpart to the Edgeworth-Kuiper belt, and that objects from the belt can enter the disk as a result of collisions and gradual dynamical evolution. However, this hypothesis remains under debate, and with the ongoing development of models which suggest that the formation of our Solar system involved significant migration of the outer planets, it is perfectly possible that the Scattered Disk was principally formed as a result of that migration. It should be noted that there are additional populations of resonant and non-resonant objects that move on dynamically stable orbits in this region. The most notable resonant population is the Plutinos, a family of objects trapped within the Neptunian 2:3 MMR, at a semi-major axis of ~39.5 AU, many of which move on orbits so eccentric that they cross the orbit of that planet. These objects are generally not considered members of the Classical belt or the Scattered Disk, but rather are considered in much the same manner as the Trojans – objects moving on orbits that, were it not for the protective effect of the resonance in which they reside, would display significant dynamical instability, to such a degree that the population would be lost on an astronomically short timescale. Further out, a group of objects known as the Detached disk move on eccentric orbits similar to those of the Scattered Disk, but with perihelia so far from the Sun that Neptune can have no significant influence on their long term dynamical evolution (hence, they are detached from that planet's influence). For more information on the classification of this zoo of planetesimals, we direct the interested reader to Lykawka & Mukai, 2007.

## 2. The Neptune Trojans

Planetary Trojans are objects which orbit the Sun trapped within the 1:1 mean-motion resonance of a given planet. In the simplest terms, this means that such objects orbit the Sun with essentially the same orbital period as the planet, on approximately the same orbit, moving such that they are protected by the action of the resonance from ever experiencing a close encounter with that planet. Most such objects move on tadpole-shaped orbits, librating around the L4 and L5 Lagrange points, located 60° ahead and behind of the planet in its orbit. These Lagrange points offer regions of stability in which objects can remain trapped on timescales of billions of years (e.g. Holman & Wisdom 1993; Murray & Dermott 1999; Nesvorný & Dones, 2002). A few objects trapped in Trojan orbits follow less stable horseshoe-shaped paths, moving between the L4 and L5 Lagrange points, but still never approaching their host planet particularly closely. Such orbits are typically somewhat less stable than their tadpole brethren, and as such are less well represented in the catalogue of Solar system Trojans. For an aesthetically pleasing and simple illustration of both tadpole and horseshoe Trojan behaviour, we direct the interested reader to figs 1 and 2 of Chebotarev (1974).

The most famous Trojan population is that hosted by the planet Jupiter. The first member, 588 Achilles, was discovered in 1906, and there are now over 3000 such objects known. It is postulated that there may actually be more objects in the Jovian Trojan population than the asteroid belt – the only reason we have found fewer to date is simply that they are further from the Sun, and hence significantly fainter and harder to detect. For the same reason, the first Neptunian Trojan was not discovered until 2001 – again, these objects, being still more distant, are even fainter and harder to spot. However, the presence of such a population had been, to an extent, anticipated long before that discovery (e.g. Mikkola & Innanen 1992).

To date, only six Neptune Trojans have been discovered, namely 2001 QR322, 2004 UP10, 2005 TN53, 2005 TO74, 2006 RJ103 and 2007 VL305. All six objects move on tadpole orbits around the Neptunian L4 Lagrange point (for a table detailing their orbital parameters, we refer the interested reader to Lykawka et al., 2009, table 1). Despite this apparent lack of observational data, the population already displays a number of unexpected features. First, based on this small observational sample, it is estimated that the Neptune Trojan population is at least as numerous as that of the main Asteroid belt, and likely actually outnumbers that population by an order of magnitude (Sheppard & Trujillo 2006). Furthermore, although it had been widely postulated that any Neptunian Trojan population would be dynamically cold (i.e. the members having very low orbital inclinations and eccentricities), it instead seems to be considerably more excited than was expected, as evidenced by the two moderately inclined Trojans (2005 TO74 & 2006 RJ103) and those with unexpectedly high inclination (2005 TN53 & 2007 VL305). This result seems particularly surprising when one considers that our Solar system is believed to have formed from a dynamically cold disk of debris (with particles on very low inclination and eccentricity orbits). Had the Trojans formed from such a disk, then it seems reasonable to anticipate they would also lie on very dynamically cold orbits (e.g. Chiang & Lithwick 2005; Hahn & Malhotra 2005). The situation is exacerbated by the biases inherent in the search for objects beyond Neptune. Surveys often concentrate on areas in the plane of the ecliptic, which means that objects on low inclination orbits are far more likely to be discovered than those at high inclination. Sheppard & Trujillo (2006) argue that finding so many highly inclined Trojans in the current sample suggests that there are likely to be far more Trojans on high inclination orbits than there are on dynamically cold orbits, which in turn has implications for the formation of the population itself. As such, a number of authors now view the Neptune Trojans as an exciting new test bed for models of the formation of our Solar system (e.g. Ford & Chiang 2007; Lykawka & Mukai 2008; Levison et al. 2008; Lykawka et al. 2009, 2010; Lykawka & Horner 2010). Over the coming decade a number of observational programs will greatly increase our understanding of the Neptunian Trojan population, from detailed studies of individual Trojans using the *Herschel* space telescope (e.g. Mueller et al. 2009) to the potentially vast numbers of new objects that will be discovered by surveys such as Pan-STARRS (Jewitt 2003) and the LSST (Ivezic et al. 2008). A better understanding of the structure of the Neptunian Trojan cloud will reveal a great deal about the dynamical processes which occurred during the final stages of planetary formation within our Solar system, and provide a vital extra datum for models attempting to explain the more general process through which other planetary systems form and evolve.

**3. Simulating the capture and evolution of the Neptune Trojan population**
We have carried out a number of contemporaneous studies of the Neptunian Trojan population (using the dynamical packages *MERCURY* (Chambers, 1999) and EVORB (Brunini & Melita 2002) to carry out our numerical orbital integrations), in an attempt to better understand their formation and long term evolution. That work is explained in detail elsewhere (e.g. Lykawka et al. 2009, 2010; Lykawka & Horner 2010; Horner & Lykawka 2010a, b). Here, we present a brief summary of that work, prior to detailing the key results that have implications for the flux of potential impactors to the inner Solar system.

First, to examine the stability of the current Neptune Trojan population, we carried out small scale integrations (using just a small number of test particles) of the orbits of each of the known Trojans. For these, we took the nominal orbit of each object (as of $5^{th}$ Feb 2009) and used that as the base for a population of 100 clones. These clones were placed on orbits spread evenly across the $3\sigma$ orbital uncertainties in the object's semi-major axis and eccentricity and, when combined with a test particle placed on the nominal orbit, led to a test population of 101 objects. The orbits of these particles were then integrated for a period of 10 Myr under the influence of Jupiter, Saturn, Uranus and Neptune, which allowed the various properties of their resonant behaviour to be determined (for more detail, see section 2 of Lykawka et al., 2009). Interestingly, the behaviour of the clones of 2001 QR322 suggested that it might be somewhat dynamically unstable (in contrast to the results of earlier work, e.g. Chiang et al. 2003; Marzari et al., 2003; Brasser et al. 2004; Sheppard & Trujillo 2006). This led us to a more detailed study of the dynamics of that object (Horner & Lykawka, 2010b), which followed the evolution of 19683 test particles (spread evenly across $\pm 3\sigma$ in semi-major axis, inclination and eccentricity around the nominal orbit) for a period of 1 Gyr. That study revealed that the number of clones of 2001 QR322 that remain in the Neptunian Trojan cloud decays in a roughly exponential fashion, with a typical dynamical half-life of ~550 Myr. Once the clones leave the Neptunian Trojan population, they behave as typical Centaurs, experiencing repeated close encounters with the giant planets which act to hand them back and forth, much as described in Horner et al. 2004. Indeed, a number of the clones evolved inward to become Jupiter-family comets for a period of time prior to their removal from the Solar system. The Jupiter-family comets are the component of the short-period comet population whose aphelia lie in the vicinity of Jupiter's orbit. These objects make up the great bulk of the short-period population, and typically move on orbits of period ~10 years or less. A protracted stay as a Jupiter-family comet, then, increases the likelihood of a given object having the opportunity to hit the Earth – simply, the shorter the orbital period of the object, the more potential encounters with the Earth can happen in a given period. For more details on the difference between the Jupiter-family comets and the short-period comets, we direct the interested reader to Horner et al., 2003.

In parallel to these studies, we examined the effect of planetary migration (e.g. Fernandez & Ip 1984; Malhotra 1995; Gomes et al., 2004; Hahn & Malhotra 2005) on the Neptunian Trojan population (Lykawka et al., 2009). In that work, we considered a variety of conservative scenarios detailing the final stages of planetary formation. There are two distinct ways in which an object can become a Neptune Trojan at the current time. Firstly, it is possible that they could form as such, prior to the migration of the planet, and then be carried along with it through the course of its subsequent migration to its current location. Alternatively, the object could form elsewhere, and be captured as a Neptune Trojan at a later date. Due to the various instabilities encountered by the planet through the course of its migration, it seems that the easiest way to capture such objects would be during that process, although it should be noted that Horner & Evans (2006) show that temporary capture of material to planetary Trojan clouds can happen even at the current epoch. In our work, then, we considered both possibilities. We followed the dynamical evolution of pre-formed Trojans as they were transported along with Neptune during its migration, and also examined the efficiency with which the planet captured fresh Trojans from the planetesimal disk through which it migrated.

We found that migration was typically unable to excite the pre-formed Trojan population to high inclinations and eccentricities, except when it involved the orbits of Uranus and Neptune experiencing a period of mutual excitation. In that scenario, the transport was highly inefficient, with the great majority of objects being lost (even if some were later recaptured). In the other scenarios tested, transport of Trojans was surprisingly efficient, with survival rates of up to 98%. The capture of Trojans during migration was reasonably inefficient, typically of order 0.1 – 1%. Although this sounds a small value, we note that there was likely upwards of 30 Earth-masses of material from which to capture Trojans, and so such capture is fully compatible with a large modern day Trojan population. Captured objects typically reproduced the observed spread of inclination and eccentricities of the current day Neptune Trojan population, and so seem the most promising source of that population. Again, for more detail, see Lykawka et al., 2009. We followed that work by further examination of the influence of the initial planetary architecture on the final Trojan population, finding that scenarios in which Uranus and Neptune are mutual resonant leads to a significant dynamical excitation and erosion of the Neptunian Trojan cloud (Lykawka et al., 2010).

Finally, in an ongoing project, we are examining the behaviour of the Trojan clouds produced at the end of the planetary migration runs detailed in Lykawka et al. (2009). There, we use the post-migration results as the seed for fresh integrations that follows the evolution of the Trojan clouds over the 4 Gyr since migration came to a halt. Those simulations are still ongoing, but reveal a significant flux of Trojan material onto unstable orbits in the outer Solar system.

A common theme across all these integrations is the transfer of material from theoretically stable orbits within the Neptune Trojan cloud to the dynamically unstable Centaur population. Once objects become Centaurs, they are dynamically indistinguishable from Centaurs sourced from other regions of the Solar system, and behave much as illustrated in earlier detailed studies of such objects (e.g. Horner et al., 2004a, b; di Sisto & Brunini, 2007; Volk & Malhotra, 2008; Bailey & Malhotra, 2009). Therefore, on the basis of those previous works, it seems reasonable to expect that some ~30% of escaped Neptune Trojans will, at some point, become Jupiter-family comets.

To illustrate this behaviour, we present a few examples of the evolution of objects that become Jupiter-family comets after leaving the Neptune Trojan population in figures 1 – 4. For simplicity, each object shown comes from the post-migration evolution of a population of objects obtained at the end of a scenario in which Neptune migrated slowly (taking ~50 Myr) from an initial semi-major axis of 18.1 AU to its current location. The three other giant planets (Jupiter, Saturn and Uranus) also migrated over the same timescale, such that all giant planets reached their terminal locations at the end of the period.

The post-migration evolution of the objects under the influence of Jupiter, Saturn, Uranus and Neptune was followed using *MERCURY* (Chambers, 1999) until they were removed from the Solar system (reaching an ejection distance of 50 AU)[4]. In each figure, *t=0* corresponds to the start of those integrations following the evolution of the clouds of particles once Neptune's migration has come to a halt, so the time shown on the x-axis corresponds to the post-migration evolution of the object. The objects shown in figures 1 and 2 were originally pre-formed Neptune Trojans, whilst those shown in figures 3 and 4, by contrast, were originally captured as Neptune Trojans from the planetesimal disk during the migration of the planet. Although these objects formed at different locations in the disk, they survived within the Trojan clouds for tens or hundreds of Myr after the migration of the planet ceased. It should be noted that the evolution of the objects was followed using a time-step of 0.5 years, which, though perfectly reasonable for objects in the outer Solar system, would be expected to yield errors in the orbits of objects moving with perihelia of less than ~2 or 3 AU. As such, we caution the reader to view the final evolution of the objects (once they are within the inner Solar system) as illustrative rather than definitive, especially since the terrestrial planets were not included in the integrations. Nevertheless, these examples illustrate nicely the way in which objects can move rapidly (sometimes in less than 1 Myr) from orbits in the Neptunian Trojan clouds to become short-period comets.

It is fairly straightforward to make a simple estimate of the contribution of the Neptune Trojans to the Centaur population. Here, we present two such "back of an envelope" calculations, one very conservative, the other significantly less so.

Conservatively, let us assume that the current day population of the Neptunian Trojan clouds numbers some $10^6$ objects greater than 1km in diameter. This is within the range suggested for the population of the asteroid belt at these sizes (700,000 to $1.7 \times 10^6$; Tedesco & Desert 2002). If we then assume that 2001 QR322 is particularly unusual in being dynamically unstable, and that the typical decay lifetime of objects in the Neptune Trojan population is similar to the age of the Solar system (i.e. 4 Gyr), then this suggests that, over the last 4 Gyr, a total of ~$10^6$ objects have been transferred from the Neptune Trojan population to the Centaurs (and ~300,000 of them have become Jupiter-family comets, following Horner et al. 2004a, who showed ~30% of Centaurs enter the Jupiter-family at some point). This suggests an injection of approximately one new Centaur every 4000 years. In comparison, Horner et al. (2004a) suggested that, in order to maintain a steady state Centaur population of ~44,000 objects, the Centaurs required one new member every 125 years. In other words, with these very conservative approximations, the Neptunian Trojan population is still capable of supplying ~3% of the total flux of Centaurs.

If, however, we assume that the Neptunian Trojan population is in fact of order $10^7$ objects (as suggested by the results of Sheppard and Trujillo, 2006), and that 2001 QR322 is only marginally more unstable than the typical Trojan (such that typical Trojans have a dynamical half-life of 1 Gyr), then the Trojans become the key source for the Centaurs. In this scenario, $10^7$ Trojans would have become Centaurs in the last 1 Gyr, a flux of one new object every 100 years – more than enough to maintain the predicted Centaur population without recourse to any other source region!

It is clear that significantly more study is needed of the Neptune Trojans. As new objects are discovered, dynamical studies will reveal whether 2001 QR322 is the exception, or the norm, and the true extent of the current day contribution of the Neptune Trojans to the Centaur population will become clear. It seems likely that the final result will lie somewhere between the two extremes detailed above, but it is clear that, at the very least, the Neptune Trojans should not be discounted as a source of objects that could hit the terrestrial planets.

---

[4] We note that, whilst it was possible for objects to collide with the four giant planets followed in our integrations, no impacts could be recorded on the terrestrial planets, since they themselves were left out of the integration. As such, we make no estimate of the likelihood that a given Neptunian Trojan will one day collide with the Earth.

## 4. Discussion and Conclusions

It is clear that the Neptune Trojan population represents a large reservoir of objects held essentially in cold storage since the giant planets migrated to their current locations. Although the bulk population is stable on Gyr timescales, its members are not absolutely dynamically stable, which results in a small but continuous trickle of objects leaving the cloud and moving onto dynamically unstable, planet crossing orbits. In other words, the Neptune Trojan population acts to continually resupply the Centaur population, which is the proximate source of the Jupiter-family comets, which contribute a significant fraction of the impact hazard to the Earth.

With even conservative assumptions, we have shown that the Neptune Trojans can supply at least a few percent of the material needed to maintain the Centaur and Jupiter-family populations at the currently observed level, and might even be the primary source of such objects. When one additionally considers that the Jovian Trojan population, itself likely larger than that in the asteroid belt, is likely undergoing an equivalent gradual shedding of material, it seems certain that resonant objects within planetary systems can be a significant, and hitherto overlooked, source of material feeding potentially hazardous orbits. Such a concept is of particular interest when it comes to the determination of impact fluxes and habitability in exoplanetary systems. Surveying such systems in the infra-red reveals the presence of dust linked to either the collisional grinding of, or out-gassing from, populations of potentially hazardous objects in those systems, and it is often considered that a large amount of dust in a given planetary system infers that that system would have a prohibitively high collisional regime for the development of life. It is certainly true that, in our own Solar system, collisional fragmentation plays a significant role in the transfer of material from the asteroid belt to the inner Solar system (the source of the Near-Earth asteroids), and such behaviour has also been invoked to explain the potential transfer of material from the Edgeworth-Kuiper belt to the Centaur population. In this work, however, we show that there exists a purely dynamical route by which material can be transferred from a stable reservoir (the Neptune Trojans) to the inner Solar system (the Jupiter family comets). Since such transfer does not require significant collisional grinding, it seems reasonable to consider that it would not be the source of a significant amount of dust within our system, and equally, would not yield detectable levels of dust in exoplanetary systems.

In other words, when one considers the likely habitability of exoplanetary systems, a lack of dust cannot, necessarily, be taken to directly infer a lack of potentially hazardous objects. Objects decaying from resonant populations (such as the Trojans) could easily take the role of "silent killers", making such systems significantly less clement than would otherwise be the case. It is therefore important that such populations are properly considered when the habitability of such systems is assessed, in order to understand the degree to which those systems could be considered habitable.

**Acknowledgements**
We are also particularly grateful to the referee of this paper, Barrie Jones, who made a number of helpful suggestions that greatly improved the sense, feel and flow of the paper. PSL and JAH gratefully acknowledge financial support awarded by the Daiwa Anglo-Japanese Foundation and the Sasakawa Foundation, which proved vital in arranging an extended research visit by JAH to Kobe University.

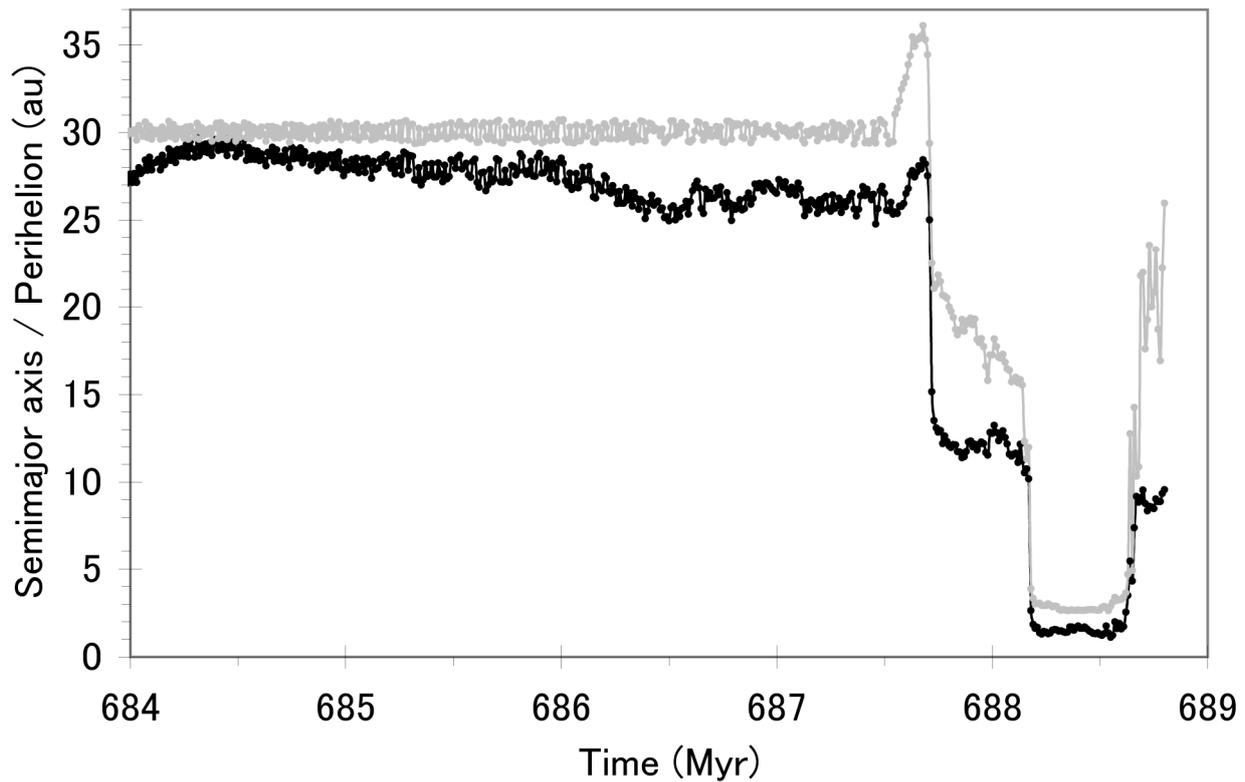

**Figure 1:** Plot The dynamical evolution of a pre-formed Neptune Trojan during the few million years around its escape from the Neptunian Trojan cloud. This object formed as a Neptune Trojan prior to Neptune's migration, and was carried by the migration of the planet from an initial semi-major axis of 18.1 AU to 30.1 AU, remaining as a Trojan throughout. Its orbit was then followed under the influence of the four giant planets, Jupiter, Saturn, Uranus and Neptune until it was ejected from the Solar system, just under 689 Myr after Neptune ceased migration. The grey line shows the evolution of the object's semi-major axis, and the black line its perihelion distance. The first 684 Myr of the objects evolution were unremarkable, but the final ~3 Myr of its life as a Trojan was marked by a gradual increase in orbital eccentricity (evidenced by the gradual inward march of the Trojan's perihelion distance. Eventually, at around 687.5 Myr, the object leaves the Neptune Trojan population, and a series of close encounters with Neptune and Uranus drive the object inward, dropping its perihelion towards the orbit of Saturn. At around 688.2 Myr, the object is captured to a short-period cometary orbit, reminiscent of that of comet 2P/Encke, where it remains for around 500 kyr, before being ejected back to the Centaur region, then removed from the Solar system just before the 689 Myr mark. The orbital elements of this Trojan at the start of the integration (t = 0) were a = 30.118392 AU, e= 0.0211010, i = 8.73000°.

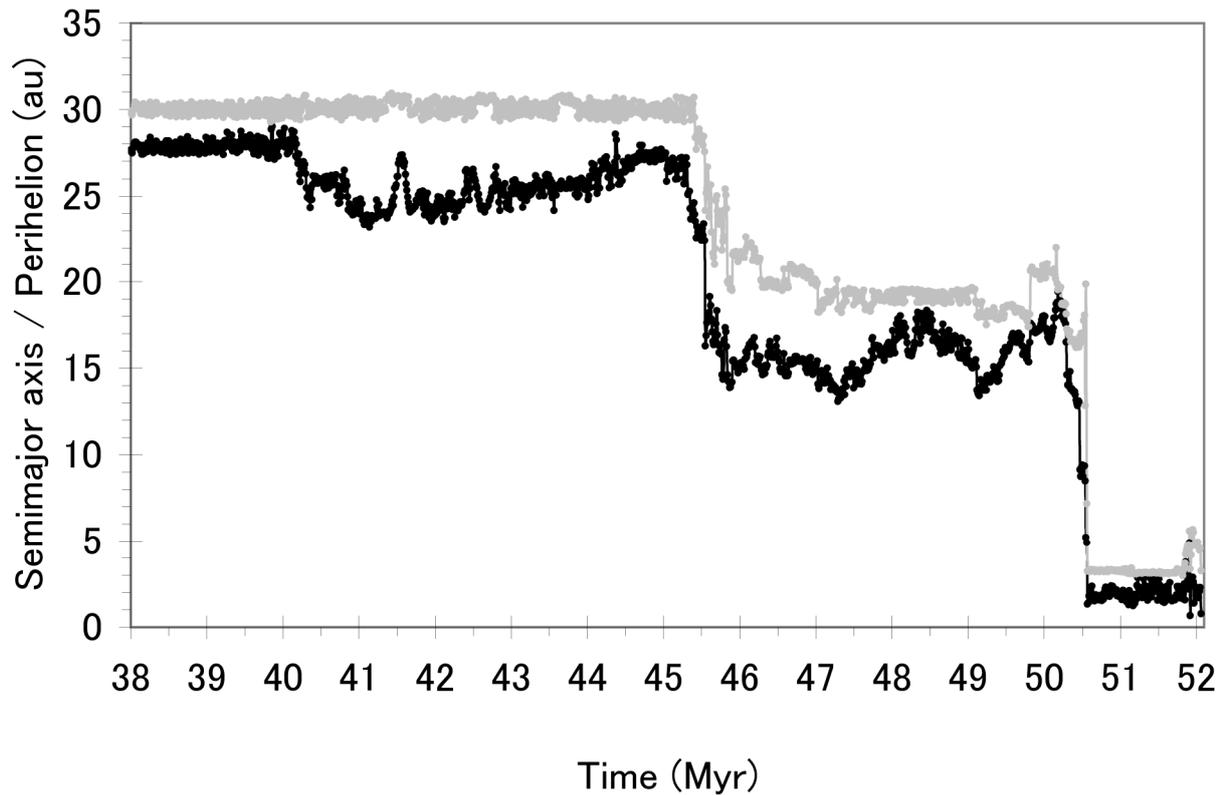

**Figure 2:** The dynamical evolution of a pre-formed Neptune Trojan during the few million years around its escape from the Neptunian Trojan cloud. This object formed as a Neptune Trojan prior to Neptune's migration, and was carried by the migration of the planet from an initial semi-major axis of 18.1 AU to 30.1 AU, remaining as a Trojan throughout. Its orbit was then followed under the influence of the four giant planets, Jupiter, Saturn, Uranus and Neptune until it was ejected from the Solar system, just over 52 Myr after Neptune ceased migration. The grey line shows the evolution of the object's semi-major axis, and the black line its perihelion distance. The first 40 Myr of the evolution of this object were unremarkable, but a sudden increase in orbital eccentricity (evidenced by a sudden drop in perihelion distance) just after the 40 Myr mark marks the beginning of its exit from the Trojan cloud. The increased eccentricity of the orbit makes it significantly less stable as a Trojan, and just 6 Myr later, the object leaves the Trojan cloud for the Centaur population. It rapidly moves inwards, ending up on an orbit with aphelion near Uranus (~19 AU), under whose control the object remains until ~50.5 Myr, when a series of close encounters with Saturn and Jupiter inject it to the inner Solar system as a short-period comet, where it remains for the last ~1.5 Myr of its life before being ejected from the Solar system by a close encounter with Jupiter. The orbital elements of this Trojan at the start of the integration (t = 0) were a = 30.042517 AU, e= 0.0355650, i = 14.80000°.

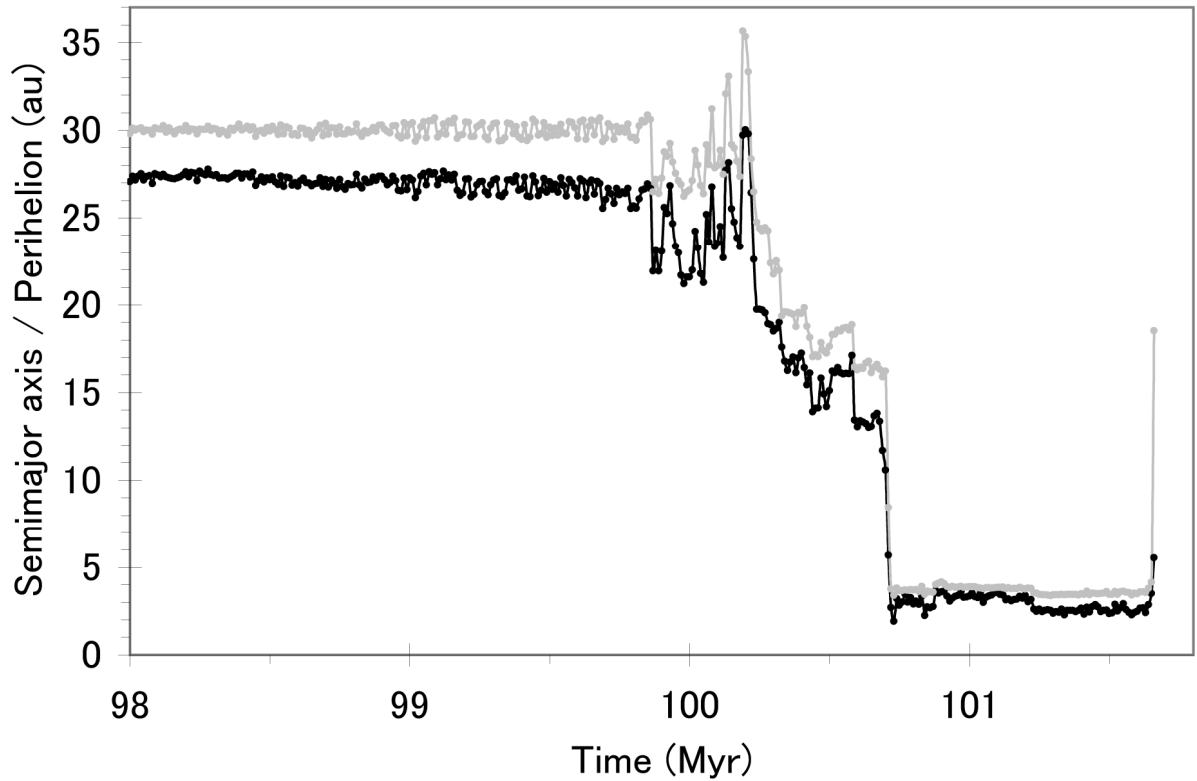

**Figure 3:** The dynamical evolution of a captured Neptune Trojan during the few million years around its escape from the Neptunian Trojan cloud. This object was initially a member of the planetesimal disk through which the planets moved as they migrated. It was captured as a Neptune Trojan during that planet's outward migration from an initial location 18.1 AU from the Sun, in the early days of our Solar system. Once Neptune ceased to migrate, having reached its current location, the dynamical evolution of the object was followed under the gravitational influence of Jupiter, Saturn, Uranus and Neptune, until it was removed from the Solar system. After almost 100 Myr evolving as a typical Neptunian Trojan, the object left the Neptune Trojan cloud, and rapidly and chaotically evolved inwards as a Centaur through a series of close encounters with Neptune, Uranus, Saturn and Jupiter. Less than 1 Myr after leaving the Trojan cloud, the object was injected to the inner Solar system on a long lived short-period cometary orbit, where it remained for almost 1 Myr before being ejected from the Solar system by a close encounter with Jupiter. The orbital elements of this Trojan at the start of the integration (t = 0) were a = 29.986243 AU, e= 0.0551690, i = 1.75000°.

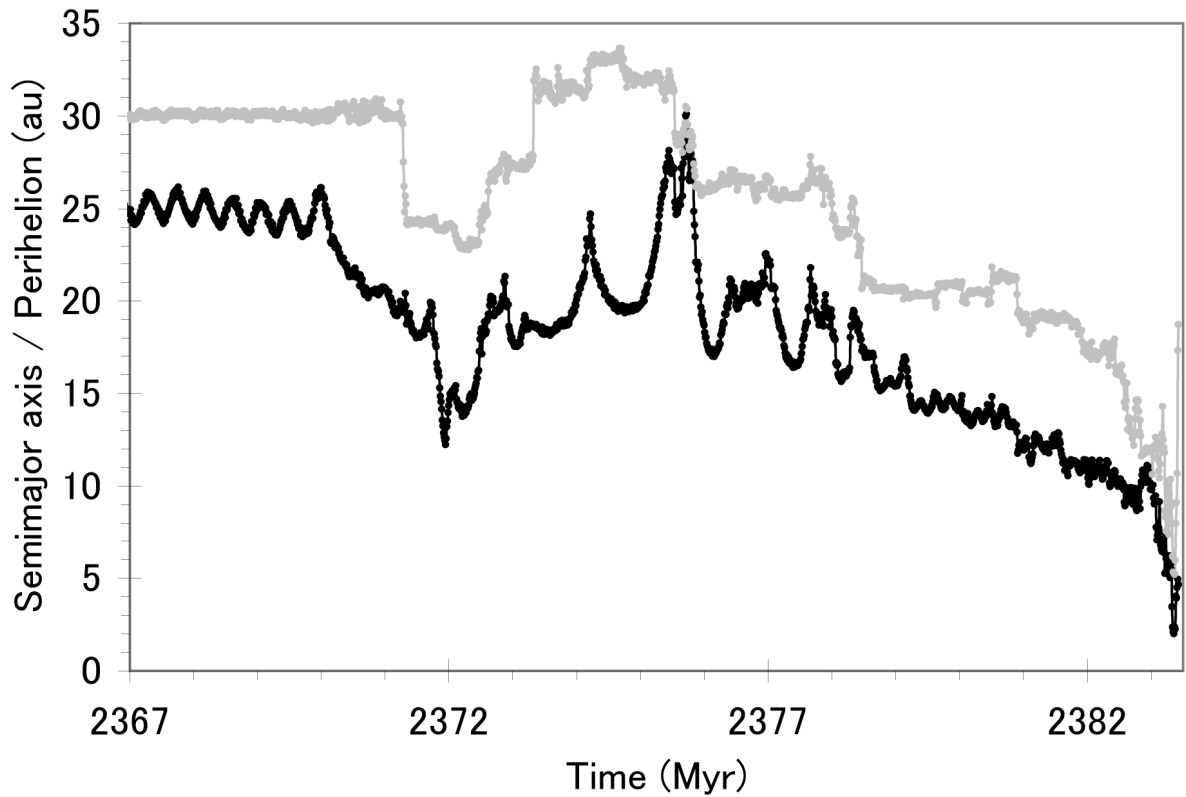

**Figure 4:** The dynamical evolution of a captured Neptune Trojan during the few million years around its escape from the Neptunian Trojan cloud. This object was initially a member of the planetesimal disk through which the planets moved as they migrated. It was captured as a Neptune Trojan during that planet's outward migration from an initial location 18.1 AU from the Sun, in the early days of our Solar system. Once Neptune ceased to migrate, having reached its current location, the dynamical evolution of the object was followed under the gravitational influence of Jupiter, Saturn, Uranus and Neptune, until it was removed from the Solar system. After about 2.370 Gyr evolving as a typical and "stable" Neptunian Trojan, the object left the Neptune Trojan cloud. It then underwent a random walk in semi-major axis and eccentricity, as a result of close encounters with Neptune and Uranus, for a period of some 12 Myr, until it finally underwent a series of close encounters with Saturn which resulted in the object being transferred to Jupiter's domain at around the 2.383 Gyr mark. In just a few hundred thousand years, the object was injected to the inner Solar system on a short lived short-period cometary orbit, upon which it remained for around 0.1 Myr before being ejected from the Solar system by a close encounter with Jupiter. The orbital evolution of this object, particularly the Gyr-timescale survival as a Neptune Trojan prior to escaping the Trojan clouds, highlights the fact that escapees from the Trojan clouds can evolve to orbits in the inner Solar system at all times through the evolution of the Solar system, even to the current day. We can therefore expect some fraction of the modern SPC population to have been sourced from the Neptunian Trojan population, so long as that population continues to decay at this epoch (a result which seems to be the case based on current observational and theoretical evidence, as described in the main text). The orbital elements of this Trojan at the start of the integration (t = 0) were a = 30.111069 AU, e= 0.1955760, i = 34.62000°.